\documentclass[12pt,english]{article}
\usepackage[T1]{fontenc}
\usepackage[latin9]{inputenc}
\usepackage[letterpaper]{geometry}
\usepackage{array}
\usepackage{amsthm}
\theoremstyle{definition}

 \usepackage{amsmath}
\usepackage{graphicx}
\usepackage{amssymb}
\usepackage{esint}
\usepackage{babel}

\makeatletter
\newcommand{\noun}[1]{\textsc{#1}}

\numberwithin{equation}{section}
\numberwithin{figure}{section}

\date{}

\makeatother

\begin{document}

\title{\noun{Nonparametric Estimation of Diffusion Coefficient}}

\title{The Implied Volatility Analysis: The South African Experience }

\author{ Romuald N. Kenmoe\footnote{:Dipartimento di Metodi Quantitativi,  University of Milano-Bicocca, Italy E-mail: r.kenmoesiyou@campus.unimib.it}
\  and \  Carine T.Tafou\footnote{Dipartimento di Scienze Bancarie, Finanziarie e Assicurative,  University of Sacred heart of Milano, Italy E-mail: tcariene1@yahoo.fr
}}
\maketitle
\section*{Abstract}

In this paper, we analyse the South African implied volatility in
various setting. We assess the information content in SAVI implied
volatility using daily markets data. Our empirical application is
focused on the FTSE/JSE Top 40 index and we emphasize our models performance
in distinct sub-periods. Our results are compared with VIX/VXN and
S\&P 500/NASDAQ 100 data in some points which are taken as our benchmark.
We find a significant negative relationship between returns and volatility,
in line with the results found in other markets. Finally, the link
between SAVI, VIX and VXN are undertaken to examine the equity market
transmission with respect to uncertainty.

\section{Introduction}

In this article, we analyze the SAVI and assess its information content
regarding realized volatility, the returns of the underlying equity
index, RiskMetric and Garch-type models using different day time horizon.
The implied volatility is the option market\textquoteright{}s forecast
of future return volatility over the remaining live of the option.
The last decades have witnessed an increase in the amount of literature
in the field of volatility forecasting and estimation. The principal
motivation consolidating this branch of research is that volatility
is the substratum in every financial application. Being that; in financial
econometrics, a great amount of research has focused on the ARCH-type
models that use past disturbances to model the variance of a time
series (see e.g. Engle, 1982 and Bollerslev, Chou, and Kroner, 2001)
for Arch type models, stochastic volatility models, (see e.g Taylor,
1986). More recently, realized volatility models is used intensively
in finance (see e.g Andersen, Bollerslev, Diebold and Ebens, 2001).
Finally there is some class of volatility based on kernel approach
of which the use is very limited in finance see Florens-Zmirou (1993),
Kenmoe and Sanfelici (2013), for an introduction and some applications.
All these models used the historical asset price series to predict
the future volatility of underlying asset An alternative way of predicting
volatility relies on exploring the information contained in the options
asset. When options price are available, one can compute the stock\textquoteright{}s
market volatility equating option price and the pricing model. As
noted by Frijns et al., (2010), this implied volatility is a forward
looking and represents the market\textquoteright{}s estimate of the
future volatility of the underlying asset for the remaining lifetime
of the option. Whaley (1993) proposes a leading work in which the
implied volatilities of various near-the-money options on the S\&P
100 index is used to construct an implied volatility index (VIX).
In 2000, the Nasdaq Volatility index (VXN) that is derived from the
implied volatility of Nasdaq-100 index (NDX) options was introduced
by CBOE. Although the principal advantage of implied volatility is
to evaluate market future volatility, it has also been used to explain
the behavior of stock market returns. In the effort of explaining
the importance of modeling and predicting asset volatility, many authors
examined the relationship between implied volatility and volatility
forecast based on historical returns in order to deliver the unbiased
and efficient forecasts of future volatility (see, e.g. Fleming et
al., 1995, Giot, 2005: Simon 2003 among others). Implied volatility
can be sued for as an underlying asset in derivatives products, the
first example in this sense is due to Deutsche Börse in 1998, with
the futures contract on the German implied volatility as the underlying
asset (VDAX), while CBOE, list futures on VIX on March 26, 2004 and
introduced options on VIX on February 24, 2006. Alternatively, implied
volatility index can serve as input variable in computing Value-At-Risk
(VaR) see Giot (2005). Finally, international integration where the
proxies are implied volatility indices spillovers has been questioned
by several authors in recent finance literature (see Siriopoulos and
Fasas, 2012, Aijö, 2008 Konstantini, Skiadopoulos and Tzagkaraki,
2008 Nikkinene and Sahlstrom, 2004). In this paper, a thorough analysis
of the South African implied Volatility (henceforth: SAVI) is undertaken.
To check our findings, we also compute some of our results on the
VIX and VXN. Therefore, these two implied volatility indices will
be considered as our benchmark. The VIX is computed and made available
by the Chicago Board of Options Exchange (CBOE) and have either found
large consensus and either used intensively both by academics and
practitioners. The VIX index use a weighted average of implied volatilities
computed from a total of eight call and put near-the money, nearby
and second nearby American options contract on the underlying S\&P
100 index by construction. This method guarantee that the index give
the implied volatility of hypothetical at-the-money option with constant
maturity of 22 trading days to expiry. The first VIX /VXN index launched
in 1993 was calculated with options based on S\&P 100 and used the
Black-Scholes/Merton model. In September 2003 the CBOE launched the
new VIX/VXN index. It used the data of S\&P 500 and based on the concept
of fair value of future variance developed by Demetrfi et al.,(1999a)
and is computed directly from market observables, which are independent
of any pricing model, such as the prices of out-of-money call and
put and interest rate, this is known as model free implied volatility.
These constraints are out of reach for emerging option markets that
are less liquid than the U.S counterpart. The SAVI was launched in
2007 by Johannesburg Stock Exchange (JSE) as an index design to measure
the market\textquoteright{}s expectation of the 3-month market volatility.
The construction is based on an more restrictive technique to VIX
method so as to respect the liquidity constraints inherent in emerging
markets. Two years later, in 2009, the SAVI index was adjusted by
the volatility skew to reflect the new way of measuring the expected
volatility. In an efficient market where options price reflect all
available information, the level of implied volatility is an indicator
of the best assessment of the expected volatility of the underlying
stock market over the remaining life of the options Giot (2005). The
majorities of empirical studies have focused on the U.S and other
developed economies where the investigations of implied volatility
have been conducted intensively. The theoretical and empirical experiments
produced in emerging markets are very limited and concentrated especially
on the Greece (see, Skiadopoulos, 2004, Siriopoulos and Fasas, 2012)
and for South-Korea Ting (2007). This limitation can be explained
by the fact that many emerging countries do not have well established
derivative markets. Contrary to this direction, the current article
contribute to the enrichment of the existing literature by focusing
our research on the evaluation of the information content in the SAVI
as the Implied volatility sentiment in the major African markets.
For this purpose, we analyze the properties of the SAVI, the Johannesburg
Stock Exchange which is the largest stock market in Africa. Although
there exist a South African Implied Volatility, there are few studies
exploring its properties and its application is rarely available in
the literature. instance Pillay and Shannon(2006) use the old methodology
of VIX, which is based on the Black and Scholes, while kotzé et al
(2009) introduced the SAVI using model-free approach. In Both article
the empirical analysis is very limited. Wandmacher and Bradfield (1998)
examined empirically the Black and Scholes assumption of a constant
volatility in the South African market. Finally, Samouilhan and Shannon
(2008) used regression and some loss functions to compare the performance
between historical volatilities (computed with GARCH-type) and implied
volatility. One question which deserves our attention is to find the
relationship between SAVI, VIX, and VXN which is known as implied
volatility spillover. There are voluminous literature on the linkages
and interactions between international stock prices and volatility.
But little had been doing concerning the transmission of implied volatilities
across the developed markets and developing markets. Gemmill and Kamiyama
(1997) examined there is a transmission across the Japanese, British,
and American markets over the period 1992-95. Skiadopoulos (2004)
investigated the linkage between Greek implied volatility, VIX, VXN.
Implied volatility propagation is of great importance to options portfolio
managers since it affects options price and hedging ratio, and it
can serve as a tool to indicate changes in expected volatility. Given
the nature of South African Stock markets in Africa, it is worth investigating
the implied volatility spillover between the leading African stock
market (which is an emergent market) and a developed market. This
study is the first step in that direction, at least in our knowledge
and it examines the properties of a measure of implied volatility
in the South African stock market. Following previous study, we assess
the efficiency, information content and absence of bias of competing
volatility forecast with respect to ex-post observed realized volatility
computed form daily returns. Furthermore, we consider the volatility
forecast based on the SAVI implied volatility index, RiskMetrics and
GJR-GARCH models and assess their relevance in efficiency, information
content and absence of bias. For each volatility forecast, we consider
various forecasting horizons. One important result of this article
is that we found an asymmetric negative relationship between SAVI
and the underlying stock index returns, this result is in line with
the previous findings.

We structured the article as follow: in section 2 the statistical
properties of SAVI and FTSE/JSE TOP 40 are presented. Section 3 scrutinizes
the relationship between implied volatility and stock market returns.
While Section 4 examines the transmission of implied volatility between
the South African and the U.S markets. In Section 5 we presents the
econometric framework and the encompassing regressions. Finally, summary
and conclusion are postponed in section 6 and 7.

\section{DATA AND DESCRIPTIVE STATISTICS}

In this section we describe the construction of the SAVI, the data
used in this article and presents some statistical properties.

\subsection{Construction of SAVI}

The South African Futures Exchange (SAFEX), launched SAVI, in 2007,
as an index designed to measure the market\textquoteright{}s expectation
of the 3-month volatility. The SAVI is based on the FTSE/JSE Top 40,
which is a capitalization index comprising the 40 most liquid stocks
trading in South African Johannesburg Stock Market(JSE) and it is
determined using at-the-money option price. The SAVI is calculated
on a daily basis, calculated the 3-month at-the money volatility and
uses bid-ask price.. Until April 2009 the SAVI was constructed using
the methodology proposed by Whaley (2000). In Mai 2009, the SAVI,
has been updated to reflect the new way of measuring the expected
3-month volatility. The new SAVI is also based on the FTSE/JSE Top
40 index, but is not only determined using the at-the-money volatilities
but using the volatility skew. This render the new index more efficient
since it incorporates a market crash protection volatility premium.
The new SAVI is calculated as the weighted average prices of calls
and puts over a wide range of strike prices, that expires in 3-month\textquoteright{}s
time. In short, 
\begin{equation}
newSAVI=\sqrt{\overset{n=F}{\underset{i=1}{\sum}}W_{iP}P_{i}(K_{i})+\overset{n=\infty}{\underset{i=n}{\sum}}W_{iC}C_{iC}(K_{i})}
\end{equation}
 where F is the current forward value of FTSE/JSE Top 40 index level,
determined using the risk-free interest rate and the dividend yield.
F marks the price boundary between the liquid puts options $P_{i}(K_{i})$
and the calls $C_{i}(K_{i})$, with strikes $K$. The price of the
call and put option are determined using the traded market volatility
skew that inspires in 3 month\textquoteright{}s time. The 3 month
volatility skew, $\sigma_{K}(0,T)$, is computed using the time weighted
interpolation function defined by 
\[
\sigma_{K}(0,T)=\sqrt{\left\{ T_{2}\sigma_{K}^{2}(0,T_{2})\left[\frac{N_{1}}{N_{2}-N_{1}}\right]+T_{1}\sigma_{K}^{2}(0,T_{1})\left[\frac{N_{1}}{N_{2}-N_{1}}\right]\right\} \frac{N_{0}}{N_{3}}}
\]
 where, $N_{0}$ is the numbers of days in the year (365 is the South
African convention), and $N_{3}$ is the number of days from the value
date to the 3 months date, $N_{1}$ and $N_{2}$ being the date of
the near skew, and nearest skew, from the 3 months skew expiry date,
respectively. The weights used in equation (1) are published by Derman
et al (1999). In this article, the period spanning from 4 May 2009
to 6 December 2012 has been set as SAVI 2, while SAVI1 starts from
the creation of the index to 23 April 2009.

\subsection{Some properties of the implied volatility}

We first present some properties of the SAVI and the FTSE/JSE Top
40 index returns. Daily data of South African stock market and SAVI
spanning from April 2007 to December 2012 are used. In figure 1, we
present time-series plot of the SAVI and FTSE/JSE TOP 40. All data
have been downloaded from Bloomberg. Negative correlation between
the underlying index and the SAVI can be guessed. Table 1 shows the
summary statistic, (viz mean, skewness, kurtosis, and the results
from first and second order autocorrelation) and the volatilities
indices. The distributional properties of return and volatilities
appear non-normal. Given that the sampling distribution for skewness
and kurtosis are normal with mean 0 and standard deviation of $\sqrt{6/T}$
and$\sqrt{24/T}$) respectively with T the
sample size, none of this distribution is well approximated by normal
distribution. The SAVI reaches his highest value on October 2008 and
the minimum was attained on December 2012. While, the FTSE/JSE Top
40 reaches its maximum value on December 2008 and its minimum value
on October 2008. The mean of SAVI is 25.77\% and the positive skewness
is indicating a longer tail and the kurtosis which exceeds the normal
distribution. Furthermore, given that the mean of daily returns is
almost zero, we can guess that there is no trend in SAVI prices. The
autocorrelation for the level of implied volatility index is positive,
indicating that the autocorrelation function decays exponentially
to zero and suggest a long-memory process. On the other hand, evidence
of negative autocorrelation in the daily returns process suggest a
mean reversing process. The significant of autocorrelation is tested
using Ljung-Box Q-Statistic. In the same Table, we summarize the statistics
of the returns of the FTSE/JSE Top 40 and the change $\triangle SAVI=SAVI_{t}-SAVI_{t-1}$
of the implied volatility indices, as well as their cross-correlations.
The sample mean for the implied volatility indices is zero indicating
the absence of a trend. The departure from normal distribution also
enhance that extreme movements in the volatility changes are more
probable than using the normal distribution. The cross-correlations
confirm the existence of leverage effect, despite the fact that it
is rather weak. The correlation between FTSE/JSE Top 40 returns and
the changes in SAVI is about -0.61 while those of the benchmark market
are -0.75 and -0.73 for S\&P 500 and NASDAQ respectively. As matter
of comparison, Skiadopoulos, 2004 find a negative correlation between
FTSE/20 and Greek implied volatility with values -0.16and -0.17. correlation
for both level and the first differences are computed and reported
in Table 3. We can see that SAVI is more correlated with VXN rather
than VIX (0.90 as opposed to 0.89). On the other hand, the correlation
of the changes of VIX (VXN) and SAVI is almost null -0.05 (-0.023).
The correlation between the changes of VIX and VXN is 0.95. The results
indicate that the implied volatility indices are correlated. This
means that South African market can reflect more information presents
in U.S markets. We can notice that the first order autocorrelation
for the \ensuremath{\Delta}SAVI is low, positive, statistically significant.
In order to investigate whether the implied volatility index time-series
are stationary, the Augmented Dickey-Fuller (ADF) unit root test at
eight lags (we use Schwarz criterion to determine the number of lags)
and without a time trend are reported in Table 2. The null hypothesis
of a unit root is rejected at 1\% significance level for the implied
volatilities. After differencing, all the series are stationary. For
accompagning the ADF test we computed the Kwiatowski-Phllips-Schmidt-Shin
(KPSS) test, both for level and first different series. This test
intended to to test the null hypothesis of stationarity (no unit root)
against the alternative non-stationarity (unit root). We reported
the results in the same table with those of ADF test. The KPSS test
results, at levels are consistent with those of the ADF the T-statistic
are significant both at 1\% and 5\% confidence level, thus rejecting
the null hypothesis of stationarity for all implied volatility indices.
Different results are obtained for the first difference of the index
series. For SAVI,VIX and VXN, the first differenced series are proved
to be stationary. This allow us to use a number of tests in our series
without imposing strict conditions.

\subsection{Day of the week effect on the SAVI}

We test the presence of the day of the week effect on the South African
implied volatility, this will allow us to check if SAVI contains any
seasonal or predictable design. We investigate the day of the week
effect by estimating the following equation, i.e., 
\begin{equation}
\triangle SAVI_{t}=\overset{5}{\underset{j=1}{\sum}}\alpha_{j}D_{jt}+\varepsilon_{t}
\end{equation}

Where j indicates the day of the week (j=1 is Monday, and so on) and
$D_{jt}$ is a dummy variable equal to one on day $j$ and zero otherwise.
$\varepsilon_{t}$ is a random term and the $\alpha_{j}$ are coefficients
to be estimated using ordinary least square (OLS). We report regression
results in Table 4. The results from the entire period revealed that,
the volatility significantly increases from Friday to Monday and insignificant
decrease can be noticed from Tuesday to Wednesday and from Thursday
to Friday. Only Monday dummy variable is significant in whole period,
with the volatility index increasing by about 8.1\%. Tuesday has the
lowest variation -0.031. In the first sub-period, which ranges from
04/05/2007 to 24/04/2009, the results confirm that Tuesday remains
the lowest volatile day, Monday always exhibit positive volatility.
In the last sub-period, spanning from 28/04/2009 to 06/04/200), Monday
dummy variable is significant and exhibits an increase of volatility
of about 10\%. We notice a trend of increase in general over the week-end,
a finding similar to that Fleming et al.(1995). However, the presence
of this seasonality is weak considering the adjusted $R^{2}$ of the
regression (-0.0006) and regression for the two sub-periods (-0.005)
and (0.0002). Significant days effects are found only for Monday,
hence it is difficult to ascertain evidence of seasonality in the
SAVI.

\section{Relationship between implied volatility stock index and underlying
stock market.}

In this section, we examine the intertemporal relationship between
FTSE/JSE Top 40 and the SAVI. Black (1976) and Christie (1982) documented
a strong negative association between stock market returns and expected
volatility observed empirically. In addition, an asymmetry relationship
between return and volatility has been evidenced in Black (1976) and
Schwert (1989,1990), that is the increase in expected volatility is
related to a given decrease in the stock market price is larger than
the respective decrease in expected volatility related to an equal
increase of the stock market price. Black (1976) motivated the asymmetry
by leverage effect; this hypothesis states that when the stock price
of a company decrease, the debt-to-equity ratio increases, leading
to higher volatility of equity returns. One other explanation known
as volatility feedback effect is due to Campbell and Hentshel (1992)
and French et al.,(1987). This hypothesis relies on the fact that,
an anticipated increase in volatility leads to an increase in the
asset\textquoteright{}s risk premium and thus in a higher expected
return, which cause current price to decline. Within the implied volatility
index literature, the evidence of a strongly asymmetric negative contemporaneous
relationship between the implied volatility index and underlying returns
is well documented. In their article for VIX construction and its
properties, Fleming et al.,(1995) examine for the first time the intertemporal
relationship between VIX and S\&P 100 returns. In the same vein, whaley
(2000), Giot (2005), among other found the presence of a statistically
significant negative contemporaneous relationship between VIX and
S\&P 500 returns. Studies dealing with other volatility indices are
consistent with the results associated to the VIX we can enumerate
among other the works of Simon (2003) and Giot (2005) for the Nasdaq
100 Volatility Index(VXN), Skiadopoulos (2004) and Siriopoulos and
Fassas (2012) for the Greek GVIX, Frijns et al., (2010) for the Australian
AVX, Gonzalez and Novales (2009) for the Spanish VIBEX-NEW. In this
article, we find a contemporaneous significant negative relationship
between variation of implied volatility (SAVI) and stock market returns.
The results are reported in Table 5.We test this relationship by following
the multivariate regression adopted in Fleming et al., (1995). We
perform a regression of the logarithmic SAVI changes on leads, lags
and contemporaneous FTSE/JSE Top 40. To assess whether there is an
asymmetric contemporaneous relationship between the SAVI and FTSE/JSE
Top 40 returns, the absolute return of lag zero has been added to
the regression. Finally, we also include a lagged value of the change
in the SAVI to check the first-order auto-correlation. The same exercise
has been computed for VIX and S\&P 500 and VXN and Nasdaq 100. We
perform the following regression:

\begin{equation}
\triangle_{h}IMP_{t}=\alpha+\beta_{-2}r_{t-2}+\beta_{-1}r_{t-1}+\beta_{0}r_{t}+\beta_{0}^{AV}\left|r_{t}\right|+\beta_{+1}r_{t+1}+\beta_{+2}r_{t+2}+\gamma\triangle_{h}IMP_{t-1}+\varepsilon_{t}
\end{equation}

where $\triangle_{h}IMP_{t}=ln(SAVI_{t})+ln(SAVI_{t-1})$

For the empirical analysis, contrarily to Fleming et al (1995), we
use the return of the SAVI rather than its level. This selection is
based on three reasons: first, both academics and practitioners are
interest in the changes or innovations of expected volatility. Second,
if stock prices follow a random walk, assessing the relationship stock
and volatility indices in levels may prove to be spurious. Finally,
implied volatility indices\textquoteright{}levels also appear to be
near random walkFleming et al (1995). From the empirical results of
Eq. (3) over the whole sample period, we notice that the coefficient
of the contemporaneous, signed change is small and negative, -0.808
with a t-statistic of -27.60 and the value of the p-value indicates
that it is significant at 10\%. The lag coefficients are negative
and significant while the lead coefficient are positive and insignificant
but much smaller in magnitude than the contemporaneous coefficient.
So, while a negative relationship exists between changes in expected
volatility and past stock returns, the opposite hold for future stock
returns. Finally, the$\beta_{0}^{AV}$ is positive and smaller than
$\beta_{0}$ ; in magnitude his value is 0.168 and the corresponding
t-statistic is 4.00. The estimate of $\beta_{0}^{AV}$ displayed in
Table 5 shows a significant asymmetry in relationship between volatility
variations and contemporaneous stock market returns. If the stock
markets return is positive, the coefficient driving the change in
volatility is $\beta_{0}^{+}=\beta_{0}+\beta_{0}^{AV}$, namely -0.6400.
A stock market increase is expected to follow a decrease in the volatility
index. On the other hand, if the stock return is negative, the coefficient
is $\beta_{0}^{-}=\beta_{0}-\beta_{0}^{AV}$, viz -0.9760. A stock
market decline is expected to follow an increase in volatility index.
The difference in the magnitudes of the coefficients, however, indicates
the asymmetry. In order to test the consistency of our findings, the
Eq. (2) is applied in the two sub-samples. In the second sub-period,
the results are consistent with the full sample results discussed
above. The significant and negative contemporaneous coefficient is
found and the $\beta_{0}^{AV}$ coefficient is lesser than the full
sample one. Results obtained from the first sub-sample are slightly
different, although the contemporaneous coefficient is still negative
and significant and coefficients are bigger than the previous ones.
Lead-lag coefficient is all negative except for $\beta_{+1}$. By
way of comparison, we regress Eq. (2) on the data of S\&P 500 and
Nasdaq 100, the regression results for VIX (VXN) and S\&P 500 (Nasdaq
100) are reported in the same table to ease the comparison. The results
are in line with the previous studies which advocates that $\beta_{0}$
have to be negative. Indeed, the results show that for the analysed
sample period there are significant negative relationship between
contemporaneous indexes returns and the corresponding volatility (the
coefficients of $\beta_{0}$ is significant with t-statistic -40.04
(-38.43) ). The coefficient $\beta_{0}^{AV}$ is positive and significant
t-statistic 5.08 (4.83). Irrespectively of the sample and indices
we find a negative $\gamma$ which reveals a negative first\textendash{}order
autocorrelation in the volatility change. The intercepts are all negative
and significant. The values of the adjusted $R^{2}$ are comprised
between 0.392 (SAVI) and 0.579 (VIX).

\section{Linkages between implied volatility}

In this section, we examine the transmission effects of implied volatility
across the CBOE and the JSE markets. Towards this end, the relationship
between VIX, VXN, and SAVI is studied. Figure 2 shows the evolution
of VIX, VXN, and SAVI over the period 2007-2012. Implied volatility
indices seem to behave similarly across markets. To analyse the linkage
between the two markets, Granger causality test and vector autoregressive
analysis (henceforth: VAR) are applied. The econometric analysis of
implied volatility crossovers could be done either in levels or in
changes. In our work, we applied the method on the first difference
thank to the stationarity of implied volatility change. In addition,
this properties of the time series are not affected by the change
of the time origin. This methods provide large information for the
analysis of the relationship between implied volatility indices and
can be useful for implementing suitable trading strategy. Preliminary
analysis of the cross\textendash{}dynamics of the implied volatility
indices and the contemporaneous and one lag cross Since the article
of Sims (1980), multivariate data analysis in the context of vector
autoregressive process become popular in econometrics. VAR models
describe the endogeneous variables relying of their own history, apart
from deterministic regressors. In this article, the following VAR(p)
system is used :

\begin{equation}
\triangle X_{t}=\alpha+\overset{n}{\underset{i=1}{\sum}}\beta_{i}\triangle X_{t-i}+\varepsilon_{t}
\end{equation}
 To make the formula clearer, we write it explicitly for the different
implied volatilities. 

\begin{equation}
\triangle VIX=\alpha^{SAVI}+\overset{n}{\underset{i=1}{\sum}}b_{i}^{SAVI}\triangle SAVI_{t-1}+\overset{n}{\underset{i=1}{\sum}}c_{i}^{VIX}\triangle VIX_{t-i1}+\overset{n}{\underset{i=1}{\sum}}d_{i}^{VXN}\triangle VXN_{t-1}+\varepsilon_{t}
\end{equation}

\begin{equation}
\triangle VIX=\alpha^{VIX}+\overset{n}{\underset{i=1}{\sum}}b_{i}^{VIX}\triangle VIX_{t-i1}+\overset{n}{\underset{i=1}{\sum}}c_{i}^{VXN}\triangle VXN_{t-1}+\overset{n}{\underset{i=1}{\sum}}d_{i}^{SAVI}\triangle SAVI_{t-1}+\varepsilon_{t}
\end{equation}

\begin{equation}
\triangle VXN=\alpha^{VXN}+\overset{n}{\underset{i=1}{\sum}}b_{i}^{VXN}\triangle VXN_{t-1}\overset{n}{+\underset{i=1}{\sum}}b_{i}^{VIX}\triangle VIX_{t-i}++\overset{n}{\underset{i=1}{\sum}}d_{i}^{SAVI}\triangle SAVI_{t-1}+\varepsilon_{t}
\end{equation}

Where, \ensuremath{\Delta}SAVI, \ensuremath{\Delta}VIX,\ensuremath{\Delta}VXN
are the daily first differences of implied volatility indices, and
are also the endogenous variables, a is a 3X1 vector of intercept,
$b_{i},$$c_{i},$ and $d_{i}$ are matrices of coefficients to be
estimated and $\varepsilon_{i}$ is a random vector of innovations
serially uncorrelated and n defines the lag order to the system. We
estimate the coefficient by means of OLS techniques. Before analysing
the Granger causality and impulse response, the optimal lag length
of the VAR(p) system has to be determined, for this end, we computed
Akaike\textquoteright{}s (AIC) and Schwatz\textquoteright{}s (SIC)
information, final prediction error,(FPE) and Lukepohl\textquoteright{}s
modified likelihood ratio (LR) test. While Akaike\textquoteright{}s
(AIC) information and final prediction error, (FPE) suggest the lag
length eight to be adapted for the VAR(p) model, Schwatz\textquoteright{}s
(SIC) information and Lukepohl\textquoteright{}s modified likelihood
ratio (LR) test suggest lag lengths of two respectively. Successively,
we have used the diagnostic test of Breitung et al.,(2004) to identify
the rightful number of lags. As the Breitung et al.,(2004) test shows
that VAR(2) specification are too restrictive, we finally opt to use
VAR(8), so the lag eight is applied in our examination. Time difference
can jeopardize our results, as the U.S underlying markets open for
trading as the continuous trading in JSE is about to stop, that is
16:00 local time.-which is also the ending time for the calculation
of the option closing prices.- Given that closing prices are used
to construct the volatility indices, \textquotedblleft{}contemporaneous\textquotedblright{}
refers to the same calendar date t even though the US and JSE indices
are measured differs. The results for the Ganger causality test are
between implied volatility changes of SAVI, VIX, VXN and reported
in Table 7. The statistics displayed are for lag order eight. A close
look of Granger test for the lags eight advocates that the Granger
VIX causes the implied volatilities of VXN and SAVI at 1\% significance
level, indicating that the expectations of the future volatilities,
as measured by the implied volatility are transmitted form the VIX
to VXN and SAVI. The results also show that implied volatility of
the VXN and SAVI are linked, however the transmission power is slightly
lower. In the Table 6, the summary of the estimations of VAR system
are displayed. The F-statistics, show that the VAR(8) model is highly
significant, as the p-values are less than 0.003. Moreover the adjusted
R-squared stand between 0.017 and 0.13.We notice that, the results
backed out by VIX and VXN are in general more robust and have the
higher coefficients. Moreover, they present the highest adjusted R-squared
which respectively 0.17 and 0.079. The results of contemporaneous
residual correlations between the markets shows that VIX and VXN are
highly correlated(0.95). We find a low, though significant correlation
between the SAVI and VIX(VXN) residuals -0.019 (-0.013). We can observe
that there is a contemporaneous spillover effect between the changes
between SAVI and both US indices. The coefficients of the lagged values
are including the intercept. We found that these do not have any additional
forecasting power. Therefore an investor can take advantage of the
information contained in the value of the changes of SAVI in order
to develop a appropriate strategy.

\section{The relation between implied and stock market volatility}

It is well accepted between academicians and practitioners that the
implied Black-Scholes volatility computed from the market options
and the model-Free counterpart are good estimates of the forward expectation
of the volatility on the underlying asset price. Therefore, in this
section we present different volatility measures that will be included
in a general regression equation. Namely, implied volatility, the
realized volatility, RiskMetrics approach and GARCH-type volatility.

\subsection{A. Implied volatility}

We first assess the information content of the implied volatility
indexes at a relatively short time-horizon (5, 10 and 22 days). In
our analysis, although mimicking relevant literature in that field,
we analyze the full sample which span from 4/5/2007 to 6/12/2012.
We proxy the measure of implied volatility by taking the level of
the SAVI, implied volatility index, for the FTSE/JSE Top 40. By definition
the forward-looking time horizon is equal to 66 trading days and the
implied volatility indices are expressed in annualized terms. We tackle
the unavailability of no implied volatility term structures using
the square root of time rule as in Giot (2005 ) and Frijns et al.
(2010) i.e to switch from a time horizon of 66 days to the required
h interval. Hence the h-day forward-looking forecast on day t is given
by the following relation: $IMP_{h,t}=\sqrt{\frac{h}{360}VOL_{t}}$
where $VOL_{t}$ is substituted by SAVI. Thus $IMP_{5,t}$ is the
expected volatility over the {[}t +1; t +5{]} period. The implied
(VOL) volatility forecasts are therefore simple re-scale version of
the VOL series.

\subsection{Realized volatility}

Given daily returns $r_{t}=ln(\frac{P_{t}}{P_{t-1}})$ for the FTSE/JSE
Top 40 index, the forward-looking realized volatility over a time
horizon of h days is computed by taking the square root of the sum
of the (future) squared returns over this h-day period. The forward-looking
realized volatility $RV_{h,t}$ at a hypothetic time t, for a generic
time period {[}t+1,t+h{]} can be computed as: 
\begin{equation}
RV_{5,t}=\sqrt{\overset{h}{\underset{j=1}{\sum}}r_{t+j}^{2}}
\end{equation}
 Note that this volatility measure is computed ex-post, i.e. at time
t+h when all returns have been observed. In this study, three values
of h are used, viz 5,10 and 22. For the encompassing regressions estimated
below, we define realized volatility computed from nonoverlapping
data. Christensen and Prabhala (1998) point out that: the use of realized
volatility computed from overlapping data in regression analysis yields
potentially big estimation problems as the regression\textquoteright{}s
residuals will be strongly auto-correlated. Therefore, the measure
of realized volatility computed using Eq. (7) and using all $RV_{h,t}$
for t = 1,\ldots{},T yields strongly correlated volatility measures.
Hence we also define realized volatility measures computed from non-overlapping
squared returns data. While Eq. (7) is still valid, we no longer compute
it for all t = 1,..,T but for a subset of those times such that the
newly defined $RV_{h,t}$ use unique data.

\subsection{RiskMetrics and GARCH-type volatility}

The last class of forecaster we consider is based on the RiskMetrics
approach. This approach can be extended to derive the well-known GARCH
(1,1). According to RiskMetrics specification, the volatility is defined
as: 
\begin{equation}
rm_{t}=(1-\lambda)r_{t-1}^{2}+\lambda rm_{t-1}
\end{equation}
 Where $\lambda$ captures the persistence in volatility and it
is set to 0.94. Note that $rm_{t}$ is an unconditional measure of
daily variance and no volatility term structure is available in this
model. The process in Eq. (8) is an interactive process and need to
be initialized at some point. The forecast we obtain from the above
equation are one day forecast. To obtain the multiple day forecast
we rescale the daily forecast as : $RM_{h,t}=\sqrt{hrm_{t}}$ Where
$RM_{h,t}$ is the h-day volatility forecast according to RiskMetrics
approach (Giot,2005) used the same modification. Finally, we construct
forecast based on ARCH type model, this type of volatility forecasting
is intensively used in finance. Under this Arch types model, the return
process is generated by $r_{t}=\mu_{t}+\varepsilon_{t}$ where $\mu_{t}$
is the conditional mean process encompassing autoregressive and moving
average terms and $\varepsilon_{t}=z_{t}\sqrt{\sigma_{t}}$ such that
$z_{t}$ is IID(0,1) and $\sigma_{t}$ is the time varying conditional
volatility process to be estimated. Symmetric GARCH The standard GARCH(1,1)
process can be modelled as 
\[
r_{t}=\varepsilon_{t},\,\varepsilon_{t}\mid\Theta\sim N(0,1)
\]
\[
\sigma_{t}^{2}=\omega+\alpha\varepsilon_{t}^{2}+\beta\sigma_{t-1}^{2}
\]
 Where $\sigma_{t}$ and $\varepsilon_{t-1}^{2}$ are the conditional
and unconditional variances of $\varepsilon_{t}^{2}$, and the long-run
variance $\omega$ is given by $\varepsilon^{2}(1-\beta-\alpha)$.
This GARCH model is symmetric in that negative and positive shocks
have the same effect on volatility $\sigma_{t}$ (time varying conditional)
volatility process to be estimated. 

Asymmetric GJR-GARCH 

In order to capture the asymmetry, GJR-GARCH pioneered by Glosten
et al.(1993) incorporates the asymmetry (or leverage effect) into
the GARCH framework by use of an indicator function. A representation
satisfying the GJR-GARCH type can be written down as follows: 
\begin{equation}
\sigma_{t}^{2}=\omega+(\alpha\varepsilon_{t}^{2}+\varphi I_{\varepsilon_{t-1<0}}\varepsilon_{t-1}^{2})+\beta\sigma_{t-1}^{2}
\end{equation}
Where the dummy variable $I_{\varepsilon_{t-1<0}}$ is such that $I_{\varepsilon_{t-1<0}}$=1
if $\varepsilon_{t-1}<0$ and $I_{\varepsilon_{t-1<0}}=0$ otherwise.
For example, in GJR-GARCH model, positive and negative shocks have
an impact of $\alpha$ and $\alpha+\varphi$, respectively. 

We can analyse the performance of the different forecaster by regressing
any competing volatility of the realized volatility using the following
formula:
\begin{equation}
RV_{k,t}=\alpha+\beta X_{k,t}+\eta_{k,t}
\end{equation}

Where $X_{k,t}$ is any of the volatility forecasters defined above.
For $X_{k,t}$ to be an unbiased forecaster of $RV_{k,t}$ we need
$\alpha=0$ and $\beta=1$. Furthermore, we evaluate the
adjusted $R^{2}$ of Eq. (5.4) to assess the predictive power of every
competing volatility forecasters. The estimates of Eq. (10) are reported
in Table 8. When we consider Implied volatility as proxied by SAVI,
we notice that it contains information regarding future realized volatility.
$\alpha$ is negative and significantly different from zero in
all cases indicating a significant bias. The $\beta$ coefficient,
is significantly different from one in many cases. Looking at the
results for the different forecast horizons reveals an interesting
pattern. The adjusted $R^{2}$ for the 5 days-ahead forecast is reasonable
high, with a value of approximately 0.50. Evidence from panel A show
that, better forecast using SAVI can be made at the 10 and 22 days
horizons. In panel B of the Table 8 we reported the results of the
estimates based on RiskMetrics approach. According to our findings,
the RiskMetrics approach does not produce unbiased forecast of the
future volatility in general, since either$\alpha$ or $\beta$
are significantly different from zero and one. Predictive power for
the RiskMetrics seem to be higher than the SAVI one. Finally, Panel
C of Table 8 reports the estimates for GJR-GARCH model. The GJR-GARCH
model performs reasonably well since it produces unbiased estimates.
The models have the highest predictive power with adjusted R\textasciicircum{}2
ranging from 0.71 to 0.96. The SAVI seems to produce worse forecasts
than RiskMetrics and GJR-GARCH model which it is the best predictor,
but when confronting with respect to the parameters estimates SAVI
perform the best. The next step of our analysis is to compare the
efficiency of the volatilities estimates to that of historic realized
volatility. For this end we estimate the following encompassing regressions
where we test the performance of one forecasting method against the
other. 

\begin{equation}
RV_{k,t}=\alpha+\beta X_{h,t}+\gamma X_{k,t}+\eta_{k,t}
\end{equation}
 Where $X_{h,t}$ and $X_{k,t}$ are the estimates based on two different
forecasting approaches. The significance of$\beta$ and $\gamma$
will indicate whether one forecasting approach dominates the other.
Alternatively, if both $\beta$ and $\gamma$ are significant
then the both forecasting approaches complement each other and the
best forecast can be made by using both forecasting approaches contemporaneously.
The estimates on Eq. (5.4) are reported in Table 9 we report the coefficients
and the statistic test in the brackets and the adjusted $R^{2}$.
In panel A, we consider the encompassing regression confronting the
RiskMetrics approach to GJR-GARCH. The coefficients are significant
for all horizons. This implies that both approaches are complementary.
The high adjusted $R^{2}$ confirmed our findings. This is furthermore
higher than the adjusted $R^{2}$ for individual regressions reported
in the Table 8. Panel B reports the encompassing regressions between
RiskMetrics approach and the SAVI. We find that in that case $\gamma$
is negative and significant. This results contrast the previous findings
available in the literature see Giot (2005), Frijns et al., (2010).
In panel C, we compare the GARCH forecast to the forecast based on
implied volatility. We notice an opposite result with confront to
RiskMetrics forecast.  $\gamma$ is positive for all horizons.
However, the GJR-GARCH contains additional information beyond the
SAVI, but this discrepancy decreases at the longer period. In all
cases we also find a highly significant SAVI series, indicating the
importance of the SAVI series.

\subsection{Discussion}

It\textquoteright{}s interesting to assess whether or not the empirical
results from econometric analysis are in line with the findings observed
in other markets. We find a statistically significant negative relationship
between the change in the level of implied volatility index and the
returns of underlying equity returns. We further find evidence for
the asymmetry in the relationship between index returns and $\triangle_{h}IMP_{t}$.
The coefficient $\beta_{0}^{AV}$ is positive and significant confirming
the existence of asymmetric relationship. The results of the sub-samples
are similar to those for the full sample. The contemporaneous coefficient
remains negative and highly significant for all sub-periods. Results
from Granger (1969) causality test suggest that the implied volatility
term structure of the VIX and VXN cause the implied volatility term
structure of the SAVI, indicating that the expectation of the future
volatilities, as measured by implied volatility term structures are
transmitted from VIX and VXN to SAVI while the reverse does not hold.
However since the U.S market is not in the same time zone, some caution
is needed when interpreting the results. We conduct a VAR test to
obtain more information about the predictive ability of the variables
and the time structures of analysis. The F-statistic show that the
VAR(8) is highly significant(very low p-value). However, we notice
that the returns of VIX are in general statistically more robust.
The results from Table 5 show the interaction of the three indices,
proving that there is, of course, no effect either on VIX or VXN from
SAVI. Finally, we notice that, the implied volatility term structure
of the VIX has an significant impact on the implied volatility term
structure of the VXN and SAVI, this is consistent for earlier findings
of linkage of the implied volatility Aijo (2008). Finally, results
in term of forecasting show that, SAVI contain important information.
On its own accord it has, the highest power relative to the competing
forecaster( GJR-GARCH and RiskMetric). When using encompassing regression
the alternative forecaster generally dominate the SAVI, but his inclusion
increases the R-adjusted. We find that RiskMetric approach move in
the oppositely with respect to the SAVI, whereas the GJR-GARCH move
in the same way and is more informative when encompassing regression
is considered.

\section{concluding remarks}

In this article, we examine the information content in the South Africa
implied volatility as measure of volatility of different settings.
The motivation of this study being that South African market is the
most important market in Africa and one lending financial market of
developing countries. We find a negative correlation between FTSE/JSE
Top 40 returns and SAVI, this is confirm by the negative and asymmetric
contemporaneous relationship between the returns of underlying asset
and the volatility which are in line with the results found in other
markets. As note by other authors the SAVI can be used as a gauge
of investor\textquoteright{}s fear. The analysing of the relationship
between SAVI and the implied proxies of the U.S markets by means of
vector autoregression (VAR) analysis and granger causality tests show
a spillover effect between the U.S markets and the South African one.
However, lead-lag effects are very weak. In general, the results suggest
that the SAVI contains important information about South African main
principal market index (FTSE/JSE Top 40) and it can provide valuable
information to market participants A large study encompassing the
other emergent markets should be helpful in shedding a light on the
behaviour of the implied volatility in these market. This topic is
currently under investigation. An analysis of the SAVI using high
frequency data can also be performed

\




\bibliographystyle{plain}
\nocite{*}

\end{document}